\documentclass[AMA,STIX1COL]{WileyNJD-v2}
\usepackage{moreverb}
\usepackage{tikz}
\usetikzlibrary{arrows,decorations.markings}

\hypersetup{
    colorlinks=true,
    linkcolor=blue,
    filecolor=magenta,      
    urlcolor=cyan,
    pdftitle={Overleaf Example},
    }

\newcommand\BibTeX{{\rmfamily B\kern-.05em \textsc{i\kern-.025em b}\kern-.08em
T\kern-.1667em\lower.7ex\hbox{E}\kern-.125emX}}

\newcommand{\be}{\begin{equation}}
\newcommand{\ee}{\end{equation}}
\newcommand{\barr}{\begin{array}}
\newcommand{\earr}{\end{array}}
\newcommand{\bea}{\begin{eqnarray}}
\newcommand{\eea}{\end{eqnarray}}
\newcommand{\beqa}{\be \begin{array}{rcl}}
\newcommand{\eeqa}{\end{array} \ee}

\newcommand{\dt}{{\cdot}}
\newcommand{\wdg}{{\wedge}}
\newcommand{\crs}{{\times}}

\newcommand{\la}{\langle}
\newcommand{\ra}{\rangle}

\newcommand{\half}{{\textstyle \frac{1}{2}}}

\newcommand{\gam}{\gamma}

\newcommand{\sig}{\sigma}

\newcommand{\da}{\partial_a}

\newcommand{\si}{\sigma_{1}}
\newcommand{\sj}{\sigma_{2}}
\newcommand{\sk}{\sigma_{3}}

\newcommand{\go}{\gamma_{0}}

\newcommand{\Rrev}{\tilde{R}}

\newcommand{\psirev}{\tilde{\psi}}

\newcommand{\phirev}{\tilde{\phi}}

\DeclareMathAlphabet{\mathbfit}{OT1}{cmr}{bx}{it}
\SetMathAlphabet\mathbfit{bold}{OT1}{cmr}{bx}{it}
\DeclareMathAlphabet{\mathbfss}{OT1}{cmss}{bx}{n}
\SetMathAlphabet\mathbfss{bold}{OT1}{cmss}{bx}{n}

\newcommand{\sut}{SU(3)}

\newcommand{\ehat}{\hat{e}}
\newcommand{\fhat}{\hat{f}}
\newcommand{\Ehat}{\hat{E}}
\newcommand{\Fhat}{\hat{F}}
\newcommand{\Jhat}{\hat{J}}

\newcommand{\alert}[1]{{\em #1}}

\graphicspath{ {./figures/} }

\articletype{Article Type}%

\received{<day> <Month>, <year>}
\revised{<day> <Month>, <year>}
\accepted{<day> <Month>, <year>}

\begin{document}

\title{Some recent results for $SU(3)$ and Octonions within the Geometric Algebra approach to the fundamental forces of nature}

\author[1,2]{Anthony Lasenby*}


\authormark{Anthony Lasenby}

\address[1]{\orgdiv{Astrophysics Group, Cavendish Laboratory}, \orgname{University of Cambridge}, \orgaddress{\country{United Kingdom}}}

\address[2]{\orgdiv{Kavli Institute for Cosmology}, \orgname{University of Cambridge}, \orgaddress{\country{United Kingdom}}}

\corres{*Anthony N. Lasenby, Astrophysics Group, Cavendish Laboratory, JJ Thomson Avenue, Cambridge, CB3 0HE, Untited Kingdom. \email{a.n.lasenby@mrao.cam.ac.uk}}

\abstract[Abstract]{Different ways of representing the group $SU(3)$ within a Geometric Algebra approach are explored. As part of this we consider characteristic multivectors for $SU(3)$, and how these are linked with decomposition of generators into commuting bivectors. The setting for this work is within a 6d Euclidean Clifford Algebra. We then go on to consider whether the fundamental forces of particle physics might arise from symmetry considerations in just the 4d geometric algebra of spacetime --- the STA. As part of this, a representation of $SU(3)$ is found wholly within the STA, involving preservation of a bivector norm. We also show how Octonions can be fully represented within the Spacetime Algebra, which we believe will be useful in making them understandable and accessible to a new community in Physics and Engineering.
The two strands of the paper are drawn together in showing how preserving the octonion norm is the same as preserving the timelike part of the Dirac current of a particle. This suggests a new model for the symmetries preserved in particle physics. Following on from work by G\"unaydin and G\"ursey on the link between quarks, and octonions, and by Furey on chains of octonionic multiplications, we show how both of these fit well within our scheme, and give some wholly STA versions of the operations involved, which in the cases considered have easily understandable equivalents in terms of 4d geometry. Links with larger groups containing $SU(3)$, such as $G_2$ and $SU(8)$, are also considered.
}

\jnlcitation{\cname{%
\author{Lasenby A.}, 
} (\cyear{2022}), 
\ctitle{Some recent results for $SU(3)$ and Octonions within the Geometric Algebra approach to the fundamental forces of nature}, \cjournal{Math Meth Appl Sci}, \cvol{2022;00:1--??}.}

\maketitle

\section{Introduction}\label{sec1}

Geometric algebra may be a very good way of understanding where the symmetries underlying the standard model of particle physics come from. It is worthwhile seeking to understand this, since it may expose an underlying geometric content which can help shape our ideas about where extensions to the standard model can be sought, and perhaps how it may be unified with gravity.

In this contribution we discuss aspects of the application of GA to the strong force, in particular the $SU(3)$ colour forces, using initially an approach stemming from the exploration of characteristic multivectors for linear transformations discussed in a companion paper in these proceedings\cite{lasenby-char-multi}. In the current paper we use a GA version of $SU(3)$ in 6 Euclidean dimensions and shows how the characteristic multivectors are linked to the commuting bivector decomposition of $SU(3)$ recently discussed by Martin Roelfs.\cite{roelfs2021geometric}

We then go on to discuss a novel approach to the transformations of $SU(3)$ which is carried out wholly within the 4 dimensions of the Spacetime Algebra (STA). This preserves the norm of bivectors, and by operating two-sidedly is able to produce a version of the commutation relations between generators which matches that of the 6d Euclidean version. This approach is exciting in showing how just the geometric entities of 4d spacetime may be enough to encode the essential aspects of the $SU(3)$ symmetries of particles. However, it can only work with one generation of particles. Thus we then look for way of broadening the set of states and operations so that more than one generation could be accommodated, ideally whilst still working within the STA. The key step here is to introduce a non-associative product between STA spinors, which we show can reproduce the properties of the {\em Octonions}.

Octonions can seem very mysterious and difficult to approach. This is due in part to the very abstract nature of their usual definition, and the fact that they do not seem to be instantiated inside algebraic structures with which we are more familiar. Thus it would be interesting to those versed in Geometric Algebra, if the octonions could be embedded in something as familiar as the Spacetime Algebra (STA), as developed by David Hestenes\cite{hestenes1966space}. We show here how we can indeed provide a faithful representation of them within the STA, and argue how the link with the {\em Dirac current} provides an interesting basis for why the STA is in fact  a natural home for them.

We then reconsider some ideas by Furey\cite{Furey:2015tqa,Furey:2018yyy}, Dixon\cite{dixon2004division}, G\"unaydin and G\"ursey\cite{gunaydin1973quark} and others, about the link between octonions and the standard model of particle physics (the `SM'), and discuss the extent to which a reformulation of these can lead to a wholly STA-based version of the standard model. By this we mean a version where all group actions and quantities can be expressed in terms of elements of the STA, and since the STA is the geometric algebra of spacetime, can therefore be viewed as intrinsically geometric in nature.

Following the earlier development which showed how the $SU(3)$ colour group could be represented by double-sided multiplication by even-grade STA elements, we then indicate how this can be interpreted in terms of single-sided multiplication by octonions. The quantity which the $SU(3)$ colour transformations were leaving invariant was the `norm' of a general STA bivector $F$, defined as $\la \go F \go F \ra$. We show here, how this can be interpreted as a sub-part of more general requirement for the preservation of the time component of the Dirac current $\psi \go \psirev$ for a general Dirac spinor $\psi$, and not just for its bivector part, which gave the $SU(3)$ action.

Both these developments lend support to the idea that eventually we will be able to represent all the symmetries of the fundamental forces entirely within the STA, and that the apparent need for extra dimensions or abstract groups comes from the complexity that is possible in transformations between states in the STA, thus giving the geometric insight that, as we said at the beginning, could be very important for future developments. As an example of this within group theory, we consider a concrete realisation of the exceptional Lie group $G_2$ within the STA, and (although more work needs to be done on this) show a link between the STA states representing chains of left octonion multiplications and the $SU(8)$ subgroup of the exceptional group $E_7$. 

This paper will assume familiarity with the Spacetime Algebra and with Geometric Algebra more widely. For some background material on both, the paper `Geometric Algebra as a unifying language for Physics and
Engineering and its use in the study of gravity'\cite{Lasenby:2016lfl} from the 6$^{th}$ AGACSE meeting, may be useful. For those who may wish to look at a video-based form of presentation, the STA together with some applications of it in electromagnetism and quantum mechanics, is discussed in the GAME2020 lecture \url{https://www.youtube.com/watch?v=m7v2IUJtC3g&t=7s}, together with some material on $SU(3)$ from later in this paper.

\section{Characteristic multivectors and SU(3)}

\label{sect:first-su3}

In a paper also in these proceedings, Lasenby \etal\cite{lasenby-char-multi} discuss the role of `characteristic multivectors' in finding the rotor that corresponds to the finite rotation from one frame of vectors to another. 

Here, where we are concerned overall with a {\em particle physics} context, we want to start by looking at how characteristic multivectors relate to the {\em infinitesimal generators} that correspond to a generalised rotation. This will enable us to make contact with recent developments in the decomposition of bivectors into sums of mutually commuting blades as recently discussed in papers by Roelfs\cite{roelfs2021geometric} and by Roelfs and de Keninck\cite{roelfs2021graded}, whilst also allowing us to lay the groundwork for our discussions of the group $SU(3)$, which will be a central theme of this paper.

We start with a brief introduction to characteristic multivectors, and their role in the Cayley-Hamilton Theorem, by giving a summary of the relevant parts of the introduction to these in \cite{lasenby-char-multi}. Please see \cite{lasenby-char-multi} for further details.

\subsection{Characteristic Multivectors}

\label{sect:char-mult}

The essential objects here are the {\em simplicial derivatives}\cite{hestenessobczyk}. If $f$ is a vector-valued linear function of a vector $a$ living in an $m$-dimensional space, $V_m$, and the output $f(a)$ lives in the same space (the simplest case) then we can define the $r$th simplicial derivative of $f$ as follows.

We let $\{a_k\}$, $k=1,\ldots,m$ be a {\em frame} for the space and $\{a^k\}$ its reciprocal frame, which is defined by the requirement $a^i \dt a_j= \delta^i_j$ for all $i$, $j$, and where $\delta^i_j$ is the Kronecker delta. We further define $b_k=f(a_k)$. Then the $r$th simplicial derivative is
\be
\partial_{(r)} f_{(r)} = \sum \left(a^{j_r} \wdg \ldots \wdg a^{j_1}\right) \left(b_{j_1} \wdg \ldots \wdg b_{j_r}\right)
\label{eqn:char-mult-def}
\ee
where the sum over the repeated indices is restricted by $0<j_1<\ldots<j_r\leq m$.

The point about these multivector quantities is that they provide {\em invariant} information about the function $f$. The invariance is in the sense that any frame $\{a_k\}$ could be chosen, and we would still get the same objects --- they are therefore in some sense `intrinsic' to the space $V_m$ and the function $f$.

\subsection{The characteristic polynomial and the Cayley-Hamilton theorem}

\label{sect:cay-ham}

We can now employ these definitions to look at the characteristic polynomial and {\em Cayley-Hamilton theorem}. These use just the scalar parts of the various simplicial derivatives. As shown in Hestenes \& Sobczyk\cite{hestenessobczyk} Section 3-2, the characteristic polynomial is
\be
C_f(\lambda)=\sum_{s=0}^{m} (-\lambda)^{m-s} \, \partial_{(s)} \!\ast\! f_{(s)}
\ee
where $\partial_{(0)} \!\ast\! f_{(0)}$ is interpreted as 1, $\lambda$ here is the scalar argument of the polynomial function, and the $\ast$ means `scalar part of' the geometric product.

If $\lambda$ is an eigenvalue of $f$, i.e.\ $f(a)=\lambda a$, then it should be a root of the characteristic polynomial, i.e.\ we will have $C_f(\lambda)=0$.
The Cayley-Hamilton theorem is then that a linear function satisfies its own characteristic equation, i.e.\ we should find
\be
\sum_{s=0}^{m} (-1)^{m-s} \, \partial_{(s)} \!\ast\! f_{(s)} f^{m-s}(a)=0
\ee
for any input vector $a$, where $f^0(a)$ is interpreted as $a$.

\subsection{Application to generators of $SU(3)$}

As discussed above, we now want to look at the characteristic multivectors for the case of the generators of the group $SU(3)$. There has been recent attention to $SU(3)$ in a Geometric Algebra context from the work by Martin Roelfs in his thesis, and he has translated some of this into a conventional {\em matrix} approach in the paper {\em `Geometric invariant decomposition of $SU(3)$'}\cite{roelfs2021geometric}. The specific GA results for an invariant commuting bivector decomposition of a generator of $SU(3)$ are given as an example in the recent paper by Roelfs and de Keninck {\em `Graded Symmetry Groups: Plane and Simple'}\cite{roelfs2021graded}.
This latter paper systematises the decomposition of rotors and bivectors, in a way which makes explicit, in higher dimensions, the methods sketched out in Hestenes \& Sobczyk\cite{hestenessobczyk}.

As a starting point on $SU(3)$ in the present paper, we want to show how the characteristic polynomial necessary to do the split of generators into commuting bivectors, arises in the characteristic multivector approach being pursued here.

So we now let the linear function $f(a)$ discussed in Sections~\ref{sect:char-mult} and \ref{sect:cay-ham}  be not a finite \alert{rotation}
\be
a \mapsto f(a)=R a \Rrev
\ee
as considered in \cite{lasenby-char-multi}, but the \alert{generator} of a finite rotation:
\be
a \mapsto f(a)=B\dt a
\ee
for the bivector $B$, which is related to $R$ by
\be
R=e^{-B/2}
\ee

We are going to work here (though not in the second part of the paper, where we will use quite different approaches) within the approach to unitary groups of the Doran et al.\ paper {\em `Lie Groups as Spin Groups'}\cite{doran1993lie}.

The essential point is that for $SU(3)$ we end up with a 6-dimensional algebra in which the action of a bivector generator $B$ on a vector $a$ is $B\dt a$, as above.
The bivector $B$ is not the most general possible within the 6-d algebra, which has 15 independent bivectors, but is picked out by commuting with $J$, the effective imaginary for the space, which if the 6 basis vectors are $\{e_1,e_2,e_3,f_1,f_2,f_3\}$, all mutually orthogonal and squaring to 1, is defined by
\be
J=e_1f_1+e_2f_2+e_3f_3
\ee

Enumerating the possibilities yields (up to normalisation) 9 bivectors, which we call:
\be
\begin{gathered}
E_{ij}= e_i e_j + f_i f_j,\quad i<j\\
F_{ij}= e_i f_j + e_j f_i,\quad i<j\\
J_i=e_i f_i, \quad i=1,2,3 \quad \text{(no sum)}
\end{gathered}
\label{eqn:generators}
\ee

Introducing the {\em commutator product} $A\crs B=\half(AB-BA)$ for two GA quantities $A$ and $B$, then the bivector generators picked out as satisfying $B\crs J=0$ define the unitary group. Those satisfying the additional constraint that $B\dt J=0$ then define the special unitary group (for which the generator matrices are traceless).

This condition reduces the 9 generators of equation (\ref{eqn:generators}) to 8 since each $J_i$ satisfies $J_i\dt J=-1$ meaning that only combinations of the $J_i$ of the form
\be
\alpha_1 J_1 + \alpha_2 J_2 +\alpha_3 J_3, \quad \text{where} \quad \alpha_1  + \alpha_2  +\alpha_3 = 0
\ee
can be used as a generator of $SU(3)$

So far this is all well understood from the perspective of the `Lie Groups as Spin groups' paper\cite{doran1993lie} and Chapter~11 of the Doran \& Lasenby book\cite{d2003geometric}. However, we now want to give some specific details for $SU(3)$ which go beyond the details presented in those, and then discuss the role of characteristic multivectors. We note that the approach being followed here as regards how the `Lie Groups as Spin groups' setup is implemented, differs somewhat from the Roelfs approach, in that here, as already described, we are taking the bivectors $B$ as operating upon a concrete vector $a$ in the 6d space. In the Roelfs approach, the bivector algebra is taken as being an abstract one, basically equivalent to the matrix algebra, where we do not need to think of it operating upon anything. Given that our $B$ operates via a `dot product' with a vector $a$, it is not immediately obvious that this action can be stripped off from $a$ itself, but the detailed correspondence of results (see below), plus a more detailed study of the approach (unpublished notes) suggests this is not a problem.

We can relate our generators to the Gell-Mann $SU(3)$ matrices, $\lambda_1, \ldots \lambda_8$ as follows:
\be
\begin{gathered}
\lambda_2=\half E_{12}, \quad \lambda_5=\half E_{13}, \quad \lambda_7=\half E_{23},\\
\lambda_1=\half F_{12}, \quad \lambda_4=\half F_{13}, \quad \lambda_6=\half F_{23},\\
\lambda_3=\half\left(J_1-J_2\right), \quad \lambda_8=\textstyle{\frac{\sqrt{3}}{6}}\left(J_1+J_2-2J_3\right)
\label{eqn:lambda-defs}
\end{gathered}
\ee
We have used equals signs here --- the important point is that the GA quantities on the left satisfy the same algebra as the Gell-Mann matrices.
Because they are GA quantities, however, interesting geometric structural relations come to light. E.g.\ an interesting relation is
\be
I\left(\lambda_i \wdg \lambda_j\right) = \textstyle{\frac{1}{6}} \delta_{ij} J - d_{ijk}\lambda_k
\ee
where $I$ is the pseudoscalar for the overall 6d space, and $d_{ijk}$ are the \alert{symmetric} structure constants of the $SU(3)$ algebra. We can see how \alert{duality} brings the exterior product of the $\lambda$'s back to representation in terms of bivectors.

We now get to the point we wish to emphasise here w.r.t.\ characteristic multivectors and the Cayley-Hamilton theorem. With $f(a)=B\dt a$, where $B$ is a generator bivector in $SU(3)$, we will now systematically find the full set of characteristic multivectors. The first is
\be
\partial_{(1)} f_{(1)} = \sum_{j=1}^m a^j b_j=a^j(B\dt a_j) =-a^j(a_j\dt B) = -\da(a\dt B)=-2B
\ee
Note the scalar part of this is the \alert{trace}, which is 0

For the next one we define some coefficients $B_i$ via
\be
B=B_i \lambda_i
\label{eqn:gen-su3-B}
\ee
where $i$ is summed over $1,\ldots,8$, and we will define the `length' of $B$ via
\be
|B|^2=B_i B_i
\ee
Then we get
\be
\partial_{(2)} f_{(2)} = 2B^2+\textstyle{\frac{3}{2}}|B|^2 = 2 B\wdg B +\half |B|^2
\ee
and the scalar invariant is the `length' of $B$, which makes sense

For the third one we have to introduce a bit more notation. This is in order to discuss the `determinant' of the transformation.  It is useful in the current context to define exactly what determinant we are talking about via the matrices. Thus let $\Lambda_i$, $i=1,\ldots ,8$ be the actual Gell-Mann matrices (ordinarily called $\lambda_i$!), and let us form, in matrix terms, the general generator
\be
M=B_i \Lambda_i = \left[\begin{array}{ccc}
B_3+\frac{B_8\sqrt{3}}{3} & B_1-B_2 j & B_4-B_5 j \\
B_1+B_2 j & -B_3+\frac{B_8 \sqrt{3}}{3} & B_6-B_7 j \\
B_4+B_5 j & B_6+B_7 j & -\frac{2 B_8 \sqrt{3}}{3}
\end{array}\right]
\ee
The determinant we are talking about is then $\det(M)$, which is real, so we can use it without having to worry about incorporating a scalar imaginary into our 6d formalism.

So for $r=3$ we find:
\be
\partial_{(3)} f_{(3)} = -\half |B|^2 B+\textstyle{\frac{1}{4}}\left(J+4I\right) \det(M)
\label{eqn:simp-deriv3}
\ee
which has grades 2 and 6.

For $r=4$ we find:
\be
\partial_{(4)} f_{(4)} = \half \det(M) IB +\textstyle{\frac{1}{16}}|B|^4
\ee
which has grades 0 and 4.

For $r=5$
\be
\partial_{(5)} f_{(5)} = \textstyle{\frac{1}{8}} \det(M) I(B\wdg B)
\ee
which is grade 2 only.

Finally, for $r=6$ we get:
\be
\partial_{(6)} f_{(6)} = \textstyle{\frac{1}{64}} \det^2(M)
\ee
i.e.\ grade 0 only. It is interesting that this `top level' characteristic multivector, which would normally return the determinant of the transformation (see Hestenes \& Sobczyk), here returns the determinant {\em squared}. We will shortly understand why this is.

\subsection{The characteristic polynomial and Cayley-Hamilton theorem for $SU(3)$}

We can now employ these results to look at the characteristic polynomial and Cayley-Hamilton theorem. For the characteristic polynomial
\be
C_f(\lambda)=\sum_{s=0}^{m} (-\lambda)^{m-s} \, \partial_{(s)} \!\ast\! f_{(s)}
\ee
then taking the scalar parts of the above results for $f(a)=B\dt a$, we find
\be
C_f(\lambda)=\textstyle{\frac{1}{64}}\det^2(M) + \textstyle{\frac{1}{16}}|B|^4 \lambda^2 +\half|B|^2\lambda^4+\lambda^6
\ee

For the Cayley-Hamilton theorem, i.e.\ that a linear function satisfies its own characteristic equation, we should find
\be
\sum_{s=0}^{m} (-1)^{m-s} \, \partial_{(s)} \!\ast\! f_{(s)} f^{m-s}(a)=0
\ee
for any input vector $a$. Here this means that $f$ itself should satisfy:
\be
\textstyle{\frac{1}{64}}\det^2(M) a + \textstyle{\frac{1}{16}}|B|^4 f^2(a) +\half|B|^2 f^4(a) + f^6(a)=0
\ee
where we remember $f(a)=B\dt a$ and $f^0(a)=a$.

For $B$ of the completely general $SU(3)$ form given in equation (\ref{eqn:gen-su3-B}), we can explicitly evaluate this equation, and we find that the l.h.s.\ vanishes, and so the Cayley-Hamilton theorem is indeed satisfied.

At this point we can start comparing with the treatment of $SU(3)$ in the Roelfs \alert{`Geometric invariant decomposition of SU(3)'} paper\cite{roelfs2021geometric}, where he is finding the decomposition of a general $SU(3)$ traceless skew-Hermitian matrix $\mathbf{B}$ into three commuting matrices $\mathbf{b}_i$, $i=1,2,3$.

A key equation there, his equation (13) is
\be
0= \frac{1}{64}\left( \det(\mathbf{B})\right)^2-\frac{1}{64}\left({\rm tr}[\mathbf{B}^2]\right)^2\lambda+\frac{1}{4}{\rm tr}[\mathbf{B}^2]\lambda^2-\lambda^3
\label{eqn:roelfs-char-poly}
\ee
from which three roots, $\lambda_i$ can be found, and which will satisfy $\mathbf{b}_i^2=\lambda_i$.

We can compare this with our
\be
C_f(\lambda)=\textstyle{\frac{1}{64}}\det^2(M) + \textstyle{\frac{1}{16}}|B|^4 \lambda^2 +\half|B|^2\lambda^4+\lambda^6
\ee

Modulo some differences in the definition of {\sl length} of $B$, and some signs (arising from Hermitian versus anti-Hermitian matrix choices), we can see that the Cayley-Hamilton theorem applied to $f(a)=a \dt B$ is yielding an equation in which the eigenvalues are {\em squared} compared to the form got via the abstract matrix approach.

We can understand this as follows. Consider the function $f^2$. Writing out this function explicitly as acting on $a$ we get
\be
f^2(a) = \left( a \dt B \right) \dt B
\ee
and this is symmetric because its \alert{curl} vanishes, indeed
\be
\da f^2(a) = \da \left( a \dt B \right) \dt B = -|B|^2
\ee
so that $\da \wdg f^2(a) =0$.

Now in the book by Hestenes \& Sobczyk\cite{hestenessobczyk}, page 82, it says that if we wish to decompose a bivector $B$ into commuting blades, then the {\em squares} of these blades can be found since they satisfy the characteristic polynomial for the function $f(a)= a\dt B$. What is not immediately clear from this statement, is that actually the characteristic polynomial is for the {\em squared} function we have just described. This is why in forming the decomposition using the roots of (\ref{eqn:roelfs-char-poly})
we find a polynomial in the \alert{squares} of our $\lambda$ quantities, and these are the coefficients needed in the decomposition of the bivector $B$.

This also shows us why the {\em square} of the matrix determinant appears in the top level characteristic multivector for this case, since again we are effectively (referring it to the abstract matrix-type approach) working with the function $f^2$ rather than $f(a)=a\dt B$. 

All these considerations will presumably apply more generally when dealing with other groups in which we want to carry out a commuting bivector decomposition of the generators, and to compare this with what one finds via looking at the characteristic multivectors, the full power of which has probably not yet been tapped.

Before moving onto a more detailed study of $SU(3)$ in the context of a GA approach to the strong force, we note that decomposition of $SU(3)$ elements into commuting blades will come up in the perhaps surprising context of Octonions towards the end of the paper, and we will return there to an interpretation of the decomposition found in \cite{roelfs2021geometric}.

\section{GA and the forces of nature}

As expressed in the talk at GAME2020 (\url{https://www.youtube.com/watch?v=m7v2IUJtC3g&t=7s}), a particularly interesting question for someone working with GA in Physics, is whether the Spacetime Algebra (STA), is sufficient for representing and deriving {\em all} the forces of nature.

It is clear that electromagnetism and gravity fit well into this STA context (see e.g.\ the paper \alert{Gravity, gauge theories and geometric algebra}\cite{1998RSPTA.356..487L}) but what about electroweak and strong forces?

For electroweak, David Hestenes\cite{1982FoPh...12..153H}, Antony Lewis (in unpublished notes) and Chris Doran and myself\cite{d2003geometric}, have pointed out that given an STA Dirac spinor $\psi$ (which for electroweak will generally be some combination of massless left-handed electron and neutrino wavefunctions), then starting from a Lagrangian, we find that the symmetry we should look for is to find all multivectors $N$ such that when we carry out the transformation $\psi \mapsto \psi e^N$, then $\psi \go \psirev$, the {\em Dirac current}, is invariant.

This picks out the set of bivectors which commute with $\go$, i.e.\ $I\si$, $I\sj$ and $I\sk$, and the pseudoscalar $I$, which reverses to itself, but anticommutes with $\go$.
The action of the $I\si$, $I\sj$ and $I\sk$ parts is thus a `spatial rotor' $R$, which defines the $SU(2)$ part of the EW transformations, and the action of the pseudoscalar is like a `phase rotation', so is $U(1)$.

We thus get the $SU(2)\times U(1)$ gauge group of electroweak arising naturally from symmetries within the STA (which remember is the geometric algebra of real 4d spacetime --- a space we know exists!).

Mass can be generated via interaction with the Higgs `scalar', except this turns out in this STA approach to be a Pauli spinor, rather than an actual scalar. (The scalar aspect arises from the fact it does not transform under Lorentz transformations at the left of the wavefunction --- only electroweak transformations at the right.)

So electroweak fits in fine with the `STA programme', although some details remain to be worked out, and in the form just described it does not really shed light on why the $SU(2)$ symmetry only seems to be employed for left-handed wavefunctions.

\subsection{Strong forces}

\label{sect:strong-forces}

But how about \alert{Strong forces}?

We know the group we want here is $SU(3)$, but the mathematics of $SU(3)$ we were discussing just now, in either the approach given in the previous section, or Martin Roelf's version, is set in a 6d Euclidean space, in which we employ the `Lie groups as spin groups' approach in order to represent the action of unitary groups of order $N$ in a space of order $2N$. There are now two developments relative to this, which I want to talk about.

The first is earlier work (discussed in the talk at GAME2020) on how we can incorporate the strong force $SU(3)$ directly into the STA. Secondly, and linked, how it has become clear to me that \alert{Octonions} may have a very important role. Using what we show below, it is now possible to represent them just using the STA, and to show that their key feature, of preserving the norms of elements under multiplication, fits in with what both the electroweak and strong force STA approaches tell us so far.

So, to begin with this, how would we like to represent the strong force in STA? A model initially due Robert Lasenby, extends the Dirac spinor $\psi$ to be both a function of position $x$, and a linear function of a bivector argument $B$, so $\psi= \psi(B)$. The separate colour fields then arise as $\psi_i=\psi(\sigma_i)$, where $\sigma_i=\gamma_i\go$, $i=1,2,3$, are the unit norm spatial bivectors of the STA in the usual way. This means that $\psi=\psi(B,x)$, where $x$ is position, is intriguingly like a spinor version of the {\em Riemann tensor}!

Recently, I have made a definite proposal for the form of these transformations of $B$, which gives a nice picture, within the STA, of the particle physics version of the $SU(3)$ group. This is that we ask for transformations of {\em bivectors} $F$ in the STA that keep the Hermitian inner product of $F$ with itself invariant:
\be
\langle \go F \go F \rangle
\ee
(Remember the Hermitian adjoint is reversion followed by `reflection in time axis' and reversion for a bivector is just $F\mapsto -F$.)

How does this work? We do this by considering {\em double-sided} operation on $F$.
(Note a double-sided action to understand $SU(3)$ was first proposed by David Hestenes, in {\em Space-time structure of weak and electromagnetic interactions}\cite{1982FoPh...12..153H}.) 

We will not go through all the details here, but we form these from composing the single-sided actions
(for $i=1,2,3$)
\be
\begin{aligned}
\hat{e}_i &= \; \text{multiplication on the left by $\sigma_i$, so} \; \hat{e}_i(F) = \sigma_i F\\
\hat{f}_i &= \; \text{multiplication on the right by $I\sigma_i$, so} \ \hat{f}_i(F) = F I\sigma_i
\end{aligned}
\ee
and we claim the appropriate $U(n)$ generators are as follows.
\be
\begin{gathered}
\Ehat_{ij}= \ehat_i \ehat_j - \fhat_i \fhat_j,\quad i<j\\
\Fhat_{ij}= \ehat_i \fhat_j + \ehat_j \fhat_i,\quad i<j\\
\Jhat_i=\ehat_i\fhat_i, \quad i=1,2,3
\end{gathered}
\label{eqn:hats-def}
\ee
where there is no sum implied in the last line, i.e.\ each line contains three quantities, making up the expected 9 generators overall.

We restrict to $\sut$ in the same way as above. Let us look at the sum of the $\Jhat_i$,
\be
\Jhat = \Jhat_1+\Jhat_2+\Jhat_3
\ee
Since $\sigma_i F \sigma_i = -F$ (with a sum over the $i$), we have
\be
\Jhat(F)=-IF
\ee
for any bivector $F$. Thus $\Jhat$ acts like the generator of a global `phase rotation', but where the imaginary is the pseudoscalar. It is this part that is removed in making the transition to $\sut$ from $U(3)$

So as before we have to take linear combinations of the $\Jhat_i$ in which the overall sum is removed, i.e.\ the only combinations allowed are of the form
\be
\alpha_1 \Jhat_1 + \alpha_2 \Jhat_2 + \alpha_3 \Jhat_3, \quad \text{with} \quad \alpha_1+\alpha_2 +\alpha_3=0
\label{eqn:alpha-restriction}
\ee
This limits the number of independent generators to 8 instead of 9, giving the right number for $\sut$.

A really nice feature of the approach is the ease with which we can derive finite forms of the transformations. Looking at the two individual parts of the $\Ehat_{ij}$ generators, i.e.\ $\ehat_i\ehat_j$, and $\fhat_i\fhat_j$, where $i<j$, it is clear that they mutually commute, and each squares to $-1$.

Thus when we exponentiate $\Ehat_{ij}$ to obtain a finite transformation, we can immediately write (with $\alpha$ a scalar)
\be
\begin{aligned}
\exp\left(\frac{\alpha}{2} \Ehat_{ij}\right)(F)&=\exp\left(\frac{\alpha}{2}\ehat_i\ehat_j\right)\left(\exp\left(\frac{-\alpha}{2}\fhat_i\fhat_j\right)(F)\right)\\
&=\exp\left(\frac{\alpha}{2}\sigma_i\sigma_j\right) F \exp\left(\frac{\alpha}{2}\sigma_j\sigma_i\right)\\
&=R_{ij} F \Rrev_{ij}
\end{aligned}
\ee
where $R_{ij}$ is the spatial rotor $\exp\left(\frac{\alpha}{2}\sigma_i\sigma_j\right)$, which gives rotations through angle $\alpha$ about the $\epsilon_{ijk} \sigma_k$ axis.

Thus if $R$ is a general spatial rotor (and so has three d.o.f.), we can see that the $\Ehat$ sector amounts to the set of spatial rotations $RF\Rrev$. For the $\Fhat_{ij}$ generators, we again have that the two parts commute, and obtain
\be
\begin{aligned}
\exp\left(\frac{\alpha}{2} \Fhat_{ij}\right)(F)&=\exp\left(\frac{\alpha}{2}\ehat_i\fhat_j\right)\left(\exp\left(\frac{\alpha}{2}\ehat_j\fhat_i\right)(F)\right)\\
&=\cos^2 \frac{\alpha}{2} F + \half I \sin\frac{\alpha}{2}\cos\frac{\alpha}{2}\left(\sigma_i F \sigma_j + \sigma_j F \sigma_i\right)\\
&+\sin^2 \frac{\alpha}{2} \sigma_i \sigma_j F \sigma_j \sigma_i
\end{aligned}
\label{eqn:Fij-action}
\ee

To complete the set, the finite form for the $\Jhat_i$ is
\be
\exp\left(\alpha\Jhat_i\right)(F)=\cos\alpha \, F + \sin\alpha \,I\,\sigma_i F \sigma_i
\ee
with of course no sum on the r.h.s. For $\Jhat$ we have
\be
\exp\left(\alpha\Jhat\right)(F)=e^{-\alpha I}F
\ee
i.e.\ a global duality transformation. Further details will be given in Lasenby (2022, in preparation).

\section{Octonions and the STA}

There are still several further aspects that are relevant for representing the $SU(3)$ aspect of strong forces, and in particular something we have not discussed here is that the $\psi=\psi(B)$ recipe for allowing the spacetime Dirac wavefunction to respond to the internal colour forces needs to be `self-dual', i.e.\ we need $\psi(IB)=I\psi(B)$, where $I$ is the spacetime pseudoscalar. Nevertheless, it seems very likely overall that what we need can be done wholly with STA entities. 
However, such an approach deals only with one generation of particles and only with $SU(3)$ and not the other groups involved in the standard model (SM). What do we need to do to enlarge the opportunities for representing particles and forces in the SM, whilst still staying in the STA?

A clue to this comes from an increase of interest over the last few years in how {\em Octonions} could be relevant to the Standard Model. Octonions are of general interest mathematically of course, but to most people seem rather mysterious. What we wish to do here is show how they can be instantiated quite simply in the STA. This has the double benefit of making them more accessible to a lot of people (e.g.\ they can then be computed with mathematically by anyone who has access to a program able to do the STA), and also provide a new approach to using octonions in the SM, as we will see.

The octonions were first introduced as a generalisation of quaternions, by Graves in 1843 and Cayley in 1845. (The first actual paper by Graves concerning them\cite{graves1845xlvi} was published shortly after Cayley's paper\cite{cayley1845xxviii}, in 1845.) They are commonly written as 8 `units', $e_0$, $e_1$, \ldots , $e_7$, with 
\be
e_0^2=1, \quad e_i^2=-1, \quad i=1,\ldots,7
\ee

Their key feature is that they form a `normed division algebra'. Any product of octonions has a norm which is the product of the individual norms, and two non-zero octonions always multiply to produce a further non-zero octonion. As is well known, the only normed division algebras are the real numbers, complex numbers, quaternions and octonions. Among these, the complex numbers are commutative and associative, the quaternions are non-commutative but still associative, while famously the octonions are neither commutative nor associative. The doubling process which can produce a higher grade algebra from a lower one (e.g.\ the complex numbers from the reals, or the quaternions from the complex numbers), is formalised in the {\em Cayley-Dickson construction}\cite{dickson1919quaternions}. If we attempt to apply this process to the octonions themselves, we get a 16 dimensional algebra called the Sedenions, but this fails to be a normed division algebra since it contains non-zero elements which when multiplied together produce 0. The Sedenions may still be useful in particle physics, however, see e.g.\ \cite{masi2021exceptional}.

We now discuss how we can represent octonions in the STA.The key is that (in the STA) the Dirac spinors $\psi$ can be divided into parts that commute or anticommute with the unit timelike vector $\gamma_0$, and we can use these two parts (the `Pauli' and `non-Pauli' parts of the Dirac spinor), in a Cayley-Dickson type doubling procedure in a way which defines the octonionic product of two Dirac spinors in terms of these sub-parts.

An STA spinor has 8 real degrees of freedom, so we are able to identify an octonion directly with a spinor. For an `octonion' $\psi$, we define
\be
\psi_+ = \half\left(\psi + \go \psi \go\right), \quad \psi_- = \half\left(\psi -\go \psi \go\right)
\ee
as the two sub-parts of $\psi$. These will correspond to the even and odd parts of the full 3d Pauli algebra, in the usual `spacetime split' correspondence between the Pauli and Dirac geometric algebras, given by multiplication by $\gamma_0$. (See the discussion concerning Fig.~\ref{fig:sta-split} below.)

Then given two octonions, $\psi$ and $\phi$, the octonionic product between them, which we will denote `$\star$', is the Dirac spinor $\theta$ given by
\be
\boxed{
\theta=\psi \star \phi = \psi_+\phi_++\phirev_-\psi_-+\phi_-\psi_++\psi_-\phirev_+
}
\label{eqn:oct-prod}
\ee

\medskip

\noindent (We have put a box around this equation, since it is a central result of this paper.) Note carefully that the four individual products on the right hand side of this equation are all usual {\em geometric products} taking place within the ordinary STA, and the spinors involved are just ordinary STA spinors, hence our claim about being able to compute everything entirely within the STA. 

From the form of this we can see that such a product is highly unlikely to be associative, and indeed in general it is not. The property it {\em does} have, however, comes from the (essentially) defining property of the octonions, already discussed, that they form a {\em normed division algebra}.

To define the norm, we need to define a \alert{conjugate} element. For us this is (using $^*$ to denote conjugation)
\be
\boxed{
\psi^*=\psirev_+-\psi_-
}
\label{eqn:conj-def}
\ee
where the tilde over the first term on the r.h.s.\ denotes the usual GA reversion.

This yields the following norm:
\be
\boxed{
||\psi|| \equiv \psi \star  \psi^* = \half\left(\go\psi\go\psirev+\psi\go\psirev\go\right) = J\dt\go
}
\label{eqn:norm-res}
\ee
where $J=\psi\go\psirev$ is the Dirac current! (We have put boxes around the last two equations, since they are also central results of this paper.)

We know from its interpretation as a probability current, that $J$ is always non-zero and future pointing as long as $\psi$ is non-zero. This, along with its scalar nature (in the sense of grades present) coincides perfectly with properties a norm should have.

The other property we need for our product and norm to be representing a normed division algebra is the fundamental relation that the norm respects multiplication, so we require
\be
||\psi\star\phi || = ||\psi|| \, ||\phi||
\ee
for \alert{all} Dirac spinors $\psi$ and $\phi$. Given our definitions so far, this is a matter of computation, and one can find that this does indeed work. Moreover, given our reinterpretation of the octonionic norm in terms of the Dirac current, we see that with $\theta=\psi \star \phi$ it corresponds to the (ordinary STA) result
\be
||\theta||=\left(\theta \go \tilde{\theta}\right)\dt\go = \left[ \left(\psi \go \tilde{\psi}\right)\dt\go \right] \left[ \left(\phi \go \tilde{\phi}\right)\dt\go \right]
\ee

(Note the square brackets are just to separate out the two scalar quantities on the r.h.s.) It is worth noting that it is the requirement for this property which means that taking just the ordinary Clifford product between spinors in e.g.\ the STA, or Euclidean $Cl(4,0)$, does not work in terms of providing a representation of octonionic multiplication. This is because in this case we can have non-zero states with zero norm, or which multiply together to give the zero state. For example, in the STA the states $\half(1+\sigma_3)$ and $\half(1-\sigma_3)$ are clearly non-zero, but multiply together, under ordinary GA multiplication, to give 0, whilst in Euclidean $Cl(4,0)$ the states $\half(1+I)$  and $\half(1-I)$ have the same problem, which as discussed in \cite{lasenby20201d} is relevant to (necessarily) false claims to have found an associative version of the Octonions\cite{christian2018quantum}.

\subsection{Explicit assignment of octonion units to STA elements}

\begin{figure}
\begin{center}    
\includegraphics[width=0.7\textwidth]{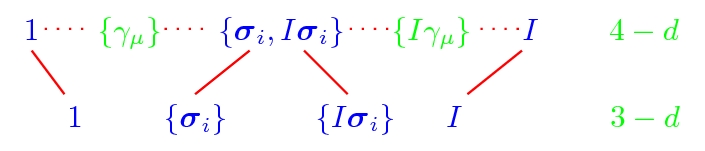}
\caption{Illustration of the relation between the even elements of the STA (blue entries on the top line) and the full 3d Pauli algebra (the geometric algebra of 3d space, bottom line).}
\label{fig:sta-split}
\end{center}
\end{figure}

We now discuss the explicit linking of the octonion units $e_0$ through $e_7$ with the 8 elements making up Dirac spinors in the STA. The latter are the even elements of the STA, which as mentioned above, can be derived as the full geometric algebra of 3d space, via the identification $\sigma_i = \gamma_i \gamma_0$, $i=1,2,3$, as illustrated in Fig.~\ref{fig:sta-split}.
Note also that this identification enables the use of the same pseudoscalar $I$ in the 4d and 3d spaces, since
\be
I=\go\gam_1\gam_2\gam_3=\sigma_1\sigma_2\sigma_3
\ee

It is fairly obvious that we will wish to associate the octonion unit $e_0$ with the scalar `1', but what about the remaining 7 entries? This depends on the multiplication table for the octonions we want to achieve, and even then there are a variety of options.

As a first example, we could aim to match the multiplication table as first given (in implied form) in the first publication concerning Octonions, i.e.\ the 1845 paper by Cayley\cite{cayley1845xxviii}. (It may be worth noting that in fact this paper was about the quite different subject of Elliptic Functions, with the Octonions occupying only a postscript. Similarly the paper by Graves, which appeared shortly afterwards\cite{graves1845xlvi}, concerned objects he had introduced called `couples', with again the Octonions only mentioned in a postscript!)

This multiplication table (the original Cayley one) is shown in Table~\ref{table:cayley-mult}
\begin{table}
\begin{center}
\begin{tabular}{|c|c|c|c|c|c|c|c|c|}
\hline & $e_{0}$ & $e_{1}$ & $e_{2}$ & $e_{3}$ & $e_{4}$ & $e_{5}$ & $e_{6}$ & $e_{7}$ \\
\hline$e_{0}$ & $e_{0}$ & $e_{1}$ & $e_{2}$ & $e_{3}$ & $e_{4}$ & $e_{5}$ & $e_{6}$ & $e_{7}$ \\
\hline$e_{1}$ & $e_{1}$ & $-e_{0}$ & $e_{3}$ & $-e_{2}$ & $e_{5}$ & $-e_{4}$ & $-e_{7}$ & $e_{6}$ \\
\hline$e_{2}$ & $e_{2}$ & $-e_{3}$ & $-e_{0}$ & $e_{1}$ & $e_{6}$ & $e_{7}$ & $-e_{4}$ & $-e_{5}$ \\
\hline$e_{3}$ & $e_{3}$ & $e_{2}$ & $-e_{1}$ & $-e_{0}$ & $e_{7}$ & $-e_{6}$ & $e_{5}$ & $-e_{4}$ \\
\hline$e_{4}$ & $e_{4}$ & $-e_{5}$ & $-e_{6}$ & $-e_{7}$ & $-e_{0}$ & $e_{1}$ & $e_{2}$ & $e_{3}$ \\
\hline$e_{5}$ & $e_{5}$ & $e_{4}$ & $-e_{7}$ & $e_{6}$ & $-e_{1}$ & $-e_{0}$ & $-e_{3}$ & $e_{2}$ \\
\hline$e_{6}$ & $e_{6}$ & $e_{7}$ & $e_{4}$ & $-e_{5}$ & $-e_{2}$ & $e_{3}$ & $-e_{0}$ & $-e_{1}$ \\
\hline$e_{7}$ & $e_{7}$ & $-e_{6}$ & $e_{5}$ & $e_{4}$ & $-e_{3}$ & $-e_{2}$ & $e_{1}$ & $-e_{0}$ \\
\hline
\end{tabular}
\caption{Multiplication table for the octonions in the form originally given by Cayley\cite{cayley1845xxviii}.}
\label{table:cayley-mult}
\end{center}
\end{table}
and we may achieve this form by the following assignment to STA elements: 
\be
e_0=1, \quad e_1=-I\sigma_1, \quad e_2=-I\sigma_2, \quad e_3=-I\sigma_3, \quad e_4=-I, \quad e_5=-\sigma_1, \quad e_6=-\sigma_2, \quad e_7=-\sigma_3
\ee
It may be wondered how assignments such as $e_5=-\sigma_1$ are going to work, since according to the multiplication table we want all the elements other than $e_0$ to have negative square, and of course $(-\sigma_1)^2=1$. This gives an opportunity to show how the product in (\ref{eqn:oct-prod}) works in an explicit case. If we put $\psi=\phi=\sigma_1$, and note $\phi_-=\sigma_1$ and $\phi_+=0$, then the product returns just
\be
\theta=\phirev_- \phi =\left(\tilde{\sig}_1\right) \sig_1 = -\sig_1 \sig_1 =-1
\label{eqn:sig1-sq}
\ee
as hoped. More generally we can see that `units' $e_i$, that commute with $\go$ will just be squared (in the STA sense), and units that anticommute with $\go$ will be multiplied by their reverse, and this pattern ensures a square in the octonion sense of $-1$ for each $i=1,\ldots,7$.  

It is good that we are able to reproduce the original multiplication laws as stated by Cayley, but this is just one table amongst a possible 480 variations all of which are compatible with the assignment $e_0=1$ (see e.g.\ the Wikipedia page \url{https://en.wikipedia.org/wiki/Octonion} for more details on this). A more recent choice of multiplication table to work from, is that given in the influential review of Octonions written in 2002 by John Baez\cite{baez2002octonions}. This table is well adapted to work with a popular way of expressing the rules of octonion multiplication, which is the {\em Fano plane}. In this, the octonion units are arranged at the vertices, midpoints and centre of an equilateral triangle, and the multiplication rules can be stated in terms of cyclic progression around the nodes.

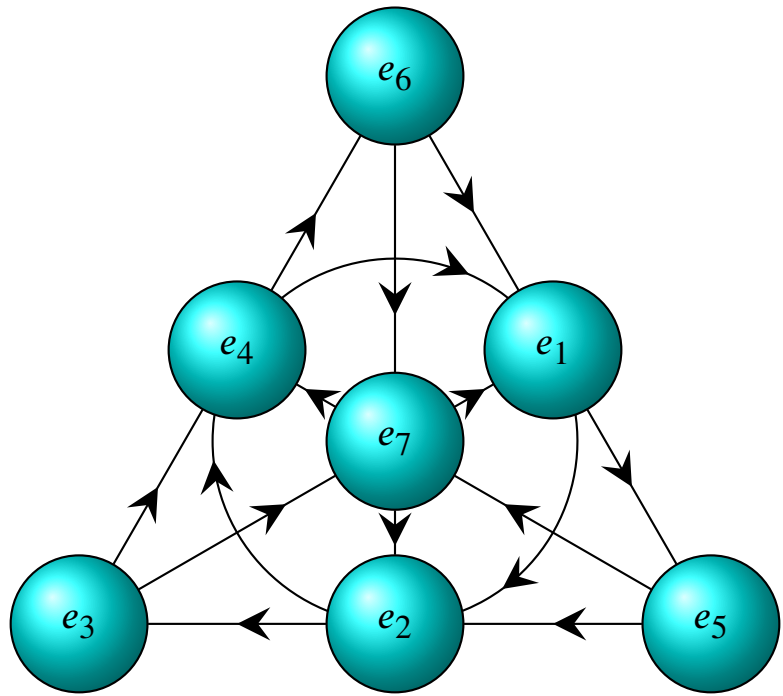
\begin{figure}
\begin{center}
\begin{tikzpicture}[thick,scale=0.6, every node/.style={transform shape}]
\tikzstyle{point}=[ball color=cyan, circle, draw=black, inner sep=0.1cm, minimum size = 3cm]
\node (v7) at (0,0) [point] {\Huge $e_7$};
\draw[thick,decoration={markings, mark=at position 0.7 with {\arrow[scale=3]{stealth}}},postaction={decorate}] (150:4cm) arc [start angle=150, end angle=30, radius=4cm];
\draw[thick,decoration={markings, mark=at position 0.7 with {\arrow[scale=3]{stealth}}},postaction={decorate}] (30:4cm) arc [start angle=30, end angle=-90, radius=4cm];
\draw[thick,decoration={markings, mark=at position 0.7 with {\arrow[scale=3]{stealth}}},postaction={decorate}] (-90:4cm) arc [start angle=-90, end angle=-210, radius=4cm];
\node (v6) at (90:8cm) [point] {\Huge $e_6$};
\node (v3) at (210:8cm) [point] {\Huge $e_3$};
\node (v5) at (330:8cm) [point] {\Huge $e_5$};
\node (v4) at (150:4cm) [point] {\Huge $e_4$};
\node (v2) at (270:4cm) [point] {\Huge $e_2$};
\node (v1) at (30:4cm) [point] {\Huge $e_1$};
\begin{scope}[thick, every node/.style={sloped,allow upside down}]
\draw[thick,decoration={markings, mark=at position 0.5 with {\arrow[scale=3]{stealth}}},postaction={decorate}] (v3) -- (v4);
\draw[thick,decoration={markings, mark=at position 0.5 with {\arrow[scale=3]{stealth}}},postaction={decorate}] (v4) -- (v6);
\draw[thick,decoration={markings, mark=at position 0.5 with {\arrow[scale=3]{stealth}}},postaction={decorate}] (v6) -- (v1);
\draw[thick,decoration={markings, mark=at position 0.5 with {\arrow[scale=3]{stealth}}},postaction={decorate}] (v1) -- (v5);
\draw[thick,decoration={markings, mark=at position 0.5 with {\arrow[scale=3]{stealth}}},postaction={decorate}] (v5) -- (v2);
\draw[thick,decoration={markings, mark=at position 0.5 with {\arrow[scale=3]{stealth}}},postaction={decorate}] (v2) -- (v3);
\draw[thick,decoration={markings, mark=at position 0.75 with {\arrow[scale=3]{stealth}}},postaction={decorate}] (v5) -- (v7);
\draw[thick,decoration={markings, mark=at position 0.8 with {\arrow[scale=3]{stealth}}},postaction={decorate}] (v7) -- (v4);
\draw[thick,decoration={markings, mark=at position 0.75 with {\arrow[scale=3]{stealth}}},postaction={decorate}] (v6) -- (v7);
\draw[thick,decoration={markings, mark=at position 0.8 with {\arrow[scale=3]{stealth}}},postaction={decorate}] (v7) -- (v2);
\draw[thick,decoration={markings, mark=at position 0.75 with {\arrow[scale=3]{stealth}}},postaction={decorate}] (v3) -- (v7);
\draw[thick,decoration={markings, mark=at position 0.8 with {\arrow[scale=3]{stealth}}},postaction={decorate}] (v7) -- (v1);
\end{scope}
\end{tikzpicture}
\caption{Fano plane for the multiplication table given in Table~\ref{table:baez-mult}. See text for a description of how the arrows relate to multiplication.}
\label{fig:fano-plane}
\end{center}
\end{figure}

The multiplication table given by Baez, which has also been used in the recent studies by Furey via the Fano plane version, is displayed in Table~\ref{table:baez-mult}.
\begin{table}
\begin{center}
\begin{tabular}{|c|c|c|c|c|c|c|c|c|}
\hline & $e_{0}$ & $e_{1}$ & $e_{2}$ & $e_{3}$ & $e_{4}$ & $e_{5}$ & $e_{6}$ & $e_{7}$ \\
\hline$e_{0}$ & $e_{0}$ & $e_{1}$ & $e_{2}$ & $e_{3}$ & $e_{4}$ & $e_{5}$ & $e_{6}$ & $e_{7}$ \\
\hline$e_{1}$ & $e_{1}$ & $-e_{0}$ & $e_{4}$ & $e_{7}$ & $-e_{2}$ & $e_{6}$ & $-e_{5}$ & $-e_{3}$ \\
\hline$e_{2}$ & $e_{2}$ & $-e_{4}$ & $-e_{0}$ & $e_{5}$ & $e_{1}$ & $-e_{3}$ & $e_{7}$ & $-e_{6}$ \\
\hline$e_{3}$ & $e_{3}$ & $-e_{7}$ & $-e_{5}$ & $-e_{0}$ & $e_{6}$ & $e_{2}$ & $-e_{4}$ & $e_{1}$ \\
\hline$e_{4}$ & $e_{4}$ & $e_{2}$ & $-e_{1}$ & $-e_{6}$ & $-e_{0}$ & $e_{7}$ & $e_{3}$ & $-e_{5}$ \\
\hline$e_{5}$ & $e_{5}$ & $-e_{6}$ & $e_{3}$ & $-e_{2}$ & $-e_{7}$ & $-e_{0}$ & $e_{1}$ & $e_{4}$ \\
\hline$e_{6}$ & $e_{6}$ & $e_{5}$ & $-e_{7}$ & $e_{4}$ & $-e_{3}$ & $-e_{1}$ & $-e_{0}$ & $e_{2}$ \\
\hline$e_{7}$ & $e_{7}$ & $e_{3}$ & $e_{6}$ & $-e_{1}$ & $e_{5}$ & $-e_{4}$ & $-e_{2}$ & $-e_{0}$ \\ 
\hline
\end{tabular}
\caption{Multiplication table for the octonions as given in Baez\cite{baez2002octonions}.}
\label{table:baez-mult}
\end{center}
\end{table}
Additionally, in Fig.~\ref{fig:fano-plane},
we show the Fano plane construction as given by Baez\cite{baez2002octonions} and Furey\cite{Furey:2015tqa}.
This construction is a mnemonic to aid in applying the octonion multiplication rules, as given in Table~\ref{table:baez-mult}. Specifically, multiplying, in the direction indicated by the arrows, any two quantities joined by an arc or a line, should give the third element along that arc or line. To give two examples, we see by following along an arc that $e_4$ applied to the left of $e_1$ should give $e_2$ and, following along a line, $e_6$ applied to the left of $e_1$ should give $e_5$. Both of these agree with Table~\ref{table:baez-mult}.

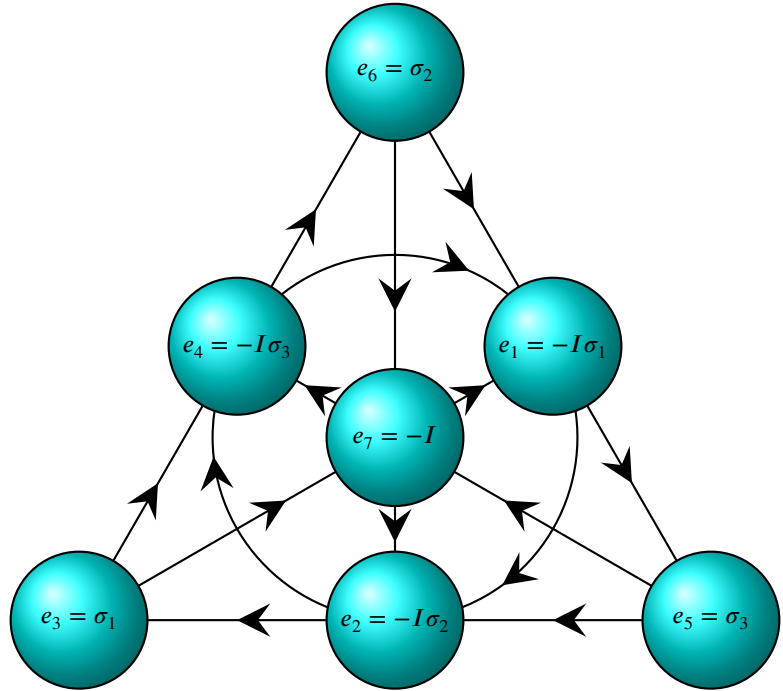
\begin{figure}
\begin{center}
\begin{tikzpicture}[thick,scale=0.6, every node/.style={transform shape}]
\tikzstyle{point}=[ball color=cyan, circle, draw=black, inner sep=0.1cm, minimum size = 3cm]
\node (v7) at (0,0) [point] {\Large$e_7=-I$};
\draw[thick,decoration={markings, mark=at position 0.7 with {\arrow[scale=3]{stealth}}},postaction={decorate}] (150:4cm) arc [start angle=150, end angle=30, radius=4cm];
\draw[thick,decoration={markings, mark=at position 0.7 with {\arrow[scale=3]{stealth}}},postaction={decorate}] (30:4cm) arc [start angle=30, end angle=-90, radius=4cm];
\draw[thick,decoration={markings, mark=at position 0.7 with {\arrow[scale=3]{stealth}}},postaction={decorate}] (-90:4cm) arc [start angle=-90, end angle=-210, radius=4cm];
\node (v6) at (90:8cm) [point] {\Large $e_6=\sigma_2$};
\node (v3) at (210:8cm) [point] {\Large $e_3=\sigma_1$};
\node (v5) at (330:8cm) [point] {\Large $e_5=\sigma_3$};
\node (v4) at (150:4cm) [point] {\Large$e_4=-I\sigma_3$};
\node (v2) at (270:4cm) [point] {\Large $e_2=-I\sigma_2$};
\node (v1) at (30:4cm) [point] {\Large $e_1=-I\sigma_1$};
\begin{scope}[thick, every node/.style={sloped,allow upside down}]
\draw[thick,decoration={markings, mark=at position 0.5 with {\arrow[scale=3]{stealth}}},postaction={decorate}] (v3) -- (v4);
\draw[thick,decoration={markings, mark=at position 0.5 with {\arrow[scale=3]{stealth}}},postaction={decorate}] (v4) -- (v6);
\draw[thick,decoration={markings, mark=at position 0.5 with {\arrow[scale=3]{stealth}}},postaction={decorate}] (v6) -- (v1);
\draw[thick,decoration={markings, mark=at position 0.5 with {\arrow[scale=3]{stealth}}},postaction={decorate}] (v1) -- (v5);
\draw[thick,decoration={markings, mark=at position 0.5 with {\arrow[scale=3]{stealth}}},postaction={decorate}] (v5) -- (v2);
\draw[thick,decoration={markings, mark=at position 0.5 with {\arrow[scale=3]{stealth}}},postaction={decorate}] (v2) -- (v3);
\draw[thick,decoration={markings, mark=at position 0.75 with {\arrow[scale=3]{stealth}}},postaction={decorate}] (v5) -- (v7);
\draw[thick,decoration={markings, mark=at position 0.8 with {\arrow[scale=3]{stealth}}},postaction={decorate}] (v7) -- (v4);
\draw[thick,decoration={markings, mark=at position 0.75 with {\arrow[scale=3]{stealth}}},postaction={decorate}] (v6) -- (v7);
\draw[thick,decoration={markings, mark=at position 0.8 with {\arrow[scale=3]{stealth}}},postaction={decorate}] (v7) -- (v2);
\draw[thick,decoration={markings, mark=at position 0.75 with {\arrow[scale=3]{stealth}}},postaction={decorate}] (v3) -- (v7);
\draw[thick,decoration={markings, mark=at position 0.8 with {\arrow[scale=3]{stealth}}},postaction={decorate}] (v7) -- (v1);
\end{scope}
\end{tikzpicture}
\caption{The same as in Fig.~\ref{fig:fano-plane} but now with the STA assignments to each node indicated.}
\label{fig:fano-plane-with-STA}
\end{center}
\end{figure}

Despite having settled on a multiplication table, we still need to make a choice of which STA elements represent which octonionic units, and we aim here to do this in such a way as the Fano plane diagram looks most symmetrical in terms of STA assignments. Given the central position of $e_7$ in the diagram, it is natural to equate this with either $I$ or $-I$, and then the sigma's and $I$sigma's fall naturally into symmetric positions around the triangle and the edge midpoints, as shown in Fig.~\ref{fig:fano-plane-with-STA}.
The resulting allocations, which we will box to indicate this is the final form we will use, are
\be
\boxed{
e_0=1, \quad e_1=-I\sigma_1, \quad e_2=-I\sigma_2, \quad e_3=\sigma_1, \quad e_4=-I\sigma_3, \quad e_5=\sigma_3, \quad e_6=\sigma_2, \quad e_7=-I
}
\label{eqn:final-STA-allocs}
\ee

We need to stress that in this diagram (Fig.~\ref{fig:fano-plane-with-STA}) the implied multiplications are {\em octonionic} multiplications as given by (\ref{eqn:oct-prod}), and therefore composed (in general) of the sum of up to 4 GA multiplications along with possible reversions. Thus there is no implication that the ordinary GA product follows the same cyclic rules as the octonion ones in this figure, though if the reader tries it, they will find that the products are correct up to signs. E.g.\ $e_6 e_1$ is equal to $e_5$ in octonion terms, meaning that $\sig_2 \star (-I\sig_1)=\sig_3$, but the GA product $\sig_2 (-I\sig_1)$ is minus rather than plus $\sig_3$.

There is more that can be said about these basic aspects of the representation of octonions within the STA, and some further examples of calculations with STA elements, based upon our main equations 
(\ref{eqn:oct-prod}), (\ref{eqn:conj-def}), (\ref{eqn:norm-res}) and
(\ref{eqn:final-STA-allocs}), can be found in the paper also in these proceedings by Eckhard Hitzer\cite{eckhard-oct}, who then goes on to consider the embedding of octonions into some different Clifford algebras, not just the STA.

\section{Octonions, the STA and particle physics}

Now, excitingly, we can start to reinterpret what we have already discussed above for the STA approach to the strong force, as a partial piece of a larger scale project for understanding the symmetries that nature may want to use. Here is a proposal:

\begin{center}
{\bf Proposal:} Nature is interested in transformations that preserve the timelike part of the STA Dirac current.
\end{center}

This is instantiated by using octonionic multiplication to represent operations upon states, since this can preserve octonionic norms, which we have said are the same thing as $J\dt\go$

This is palpably what we actually did for both the electroweak and strong force translations into the STA discussed above. For the electroweak case, we went `one further' and actually insisted the whole Dirac current was left invariant by our transformations --- this led to the $SU(2) \times U(1)$ structure.

For the strong force, then so far we can see that what we have preserved corresponds to taking just the \alert{bivector} part of $\psi$, which we wrote $F$, and demanding that
\be
\langle \go F \go F \rangle =\half\left(\go F \go F + F \go F \go\right)
\ee
be invariant --- this produced the colour forces and corresponds perfectly to what we have just been saying about preserving
\be
\half\left(\go\psi\go\psirev+\psi\go\psirev\go\right)
\ee
(note $\tilde{F}=-F$ for a bivector). Thus we must have already been doing a form of octonionic multiplication above!

Now, quite soon after quarks became established as building blocks of the Standard Model, G\"unaydin and G\"ursey\cite{gunaydin1973quark} proposed that there was a link between quark structure and the octonions. They also emphasised the role of the exceptional Lie group $G_2$ in their models, where it arises as the automorphism group of the octonions (we shall give more details about this aspect below). 

Following this, there have been many attempts to link aspects of the Standard Model to octonions, amongst them papers by Dixon (e.g.\ \cite{dixon2004division}) and Furey (e.g.\  \cite{Furey:2015tqa,Furey:2018yyy}), with these authors particularly emphasising the role of {\em chains} of octonion operators. This is a
clever technique to do with using {\em sequences} of octonions to pack in a great deal of `information' despite the relatively small size of the spaces involved. The non-associativity of the octonions is crucial to this, since then e.g.\
\begin{center}
$A(B\phi)$ for some octonions $A$ and $B$ acting on a state $\phi$
\end{center}
is {\em different} from the sequence
\begin{center}
$(AB)\phi$ also acting on $\phi$
\end{center}
The possibilities increase as we go to longer sequences, rather like a code. However, this process does not go on for ever, and as shown in the book by Conway and Smith\cite{conway2003quaternions}, 7 left multiplications or 7 right multiplications (if we build up our chains on the right instead) is the maximum we need to consider, and indeed one will be equivalent to the other (see their section 8.4, `Seven Rights can make a Left').

This means that if we consider purely left multiplications, then we should consider sequences of a maximum of 6 of these, and Furey and others (indeed it is implicit in the original G\"unaydin and G\"ursey work) show that this is isomorphic to the 6 dimensional anti-Euclidean Clifford algebra $Cl(0,6)$, i.e.\ the algebra that has all grade 1 objects squaring to $-1$, and anticommuting. That this happens we can understand quite well from the STA viewpoint. To explain this, let us look first at the example we gave above in equation (\ref{eqn:sig1-sq}) of $\sig_1 \star \sig_1$, which successfully gave $-1$.

This is {\em not} actually the type of square we are talking about here. We want to look at the action of `chains' of octonions applied at the left, so it is the application of $\sig_1$ twice which we should look at, and due to the non-associativity of the octonions, it is by no means obvious that
\be
\sig_1\star(\sig_1\star\phi) \quad \text{is equal to} \quad (\sig_1\star \sig_1) \star \phi
\ee
for a general $\phi$. Let us show, however, that this does work in this specific case, and then we will look at why it happens generally.

Let $\psi=\sigma_1$, so that $\psi_-=\sigma_1$ and $\psi_+=0$. Then using (\ref{eqn:oct-prod}), for a general Dirac spinor $\phi$, we have
\be
\sig_1 \star \phi = \phirev_-\sig_1 + \sig_1 \phirev_+
\ee
Carrying this out again, and noting now that the Pauli-even and Pauli-odd parts of $\phirev_-\sig_1 + \sig_1 \phirev_+$ are $\phirev_-\sig_1$ and $\sig_1 \phirev_+$ respectively, we get
\be
\sig_1\star(\sig_1\star\phi)=\sig_1\left(-\sig_1\phi_-\right)+\phi_+\left(-\sig_1\right)\sig_1=-\phi_--\phi_+=-\phi
\label{eqn:sig1-sq-app}
\ee
as hoped. In fact, this behaviour happens generally, since the octonions, while not associative, are what is known as an {\em alternative algebra}, which satisfies (in our STA notation)
\be
\psi\star(\psi\star\phi) =  (\psi\star \psi) \star \phi \quad \text{and} \quad (\phi\star\psi)\star \psi = \phi\star(\psi\star \psi)
\ee
for all states $\psi$ and $\phi$ in the algebra. Thus in particular, all unit elements $e_i$, $i=1,\ldots,7$ can unambiguously be described as having `square' $-1$ even when viewed in the chain sense of repeated application to the left of a state $\phi$.

In seeking to establish a correspondence with a 6d Clifford space, we also need to consider anticommutivity, and we would like to show that
\be
e_i \star \left( e_j \star \phi\right) = -e_j \star \left( e_i \star \phi\right)
\ee
for all states $\phi$, and all $i$ and $j$ from 1 to 6 with $i\neq j$. (It actually works for $i$ or $j$ equals 7 as well, but that's not needed currently.) 

The fact that the multiplication tables given above do satisfy anticommutativity, i.e.\ we have
\be
e_i \star e_j  = -e_j \star  e_i, \quad i,j=1,\ldots,7, \quad i\neq j
\ee
is not immediately relevant, due to the usual reason of non-associativity, and this time we do not have a general principle, like the alternative algebra identities, to come to our aid. However, it is fairly easy to see how this works, within the STA representation. Each of the octonion units is either Pauli-odd or Pauli-even, so we just have to consider the four possible combinations of $\theta_\pm$ and $\psi_\pm$ acting as a chain on the left of a state $\phi$.

For example, dealing with the combination which takes longest to write down, we find
\be
\theta_-\star\left(\psi_+\star\phi\right)=\psirev_+\phirev_-\theta_-+\theta_-\phirev_+\psirev_+
\ee
while
\be
\psi_+\star\left(\theta_-\star\phi\right)=\psi_+\phirev_-\theta_-+\theta_-\phirev_+\psi_+
\ee
Adding these together to form the anticommutator yields
\be
\left(\psi_++\psirev_+\right)\phirev_-\theta_-+\theta_-\phirev_+\left(\psi_++\psirev_+\right)
\ee
Now any Pauli-even non scalar quantity, i.e.\ the set $I\sig_1$, $I\sig_2$ and $I\sig_3$, is minus its own reverse, hence $\left(\psi_++\psirev_+\right)$ vanishes, as does the anticommutator, thus establishing what we wanted.

By this means, we see that $e_1$ through $e_6$, applied in chains to the left of a general state $\phi$, generate an equivalent 6d, and therefore 64 dimensional, Clifford space. It is this space, in a complexified form, which has underlain the work by Furey on linking the Standard Model with Octonions, and we see now that it is possible to represent what is happening in this space just using the STA (modulo issues about the complexification, which we return to below).

A great deal more has already been investigated and worked out by the author on this matter, and a much larger 512 dimensional space of octonionic chain states has come to light in this investigation which promises to provide a home for not just groups like $SU(3)$, $SU(2)$ and $U(1)$, which we need (in some cases in multiple copies) for the Standard Model, but for larger groups such as the exceptional group $E_8$, which has long been suspected as being highly significant in unified models of particle physics.

The details of this will be given in a separate paper (Lasenby, 2022, in preparation), and we will limit ourselves here to schematic treatments of two examples of more limited scope, and which work within just the 64 dimensional space of left-multiplication.

The first example, is of how it appears to be possible to accommodate in this space the group $SU(8)$, which we believe arises in this context as being a subgroup of the exceptional group $E_7$, which like $E_8$ has been considered several times in unified models of particle physics and string theory (e.g.\ \cite{hillmann2009generalized}). The second 
`unification', is on the use of the exceptional group $G_2$ in representing quark states. This was effectively proposed in the G\"unaydin and G\"ursey paper\cite{gunaydin1973quark}, and is in fact the first step in unification as proposed in a famous paper by Garrett Lisi, who made no connection with the octonions, but was working on the use of $E_8$ as the basic group underlying {\em all} the symmetries involved in both particle physics and gravity. The models proposed in his paper are incomplete, but the first stage, in which the much smaller exceptional group $G_2$ is used to represent both the gluons and the quarks they act upon, is fully worked out, and even though this specific group assignment is not necessarily what Nature uses, it is still of great interest to see what the approach looks like in our STA terms, and to link it with the STA approach to $SU(3)$ in Section~\ref{sect:strong-forces}.

We will start with the new approach to $SU(8)$/$E_7$, since although more complicated, the necessary states and linkages needed for the $G_2$ example become clearer if viewed in this context.

\subsection{An identification of the $SU(8)$ subgroup of $E_7$  within the 64-dimensional octonionic left multiplication space}

What we will do here is consider the full space of 64 STA states we get corresponding to the distinct chains of octonion multiplication we can carry out to the left on some initial (general) STA state $\phi$.

In seeking to interpret this in terms of a Lie algebra, we can take the operation of one state upon another as being the action of taking a {\em commutator} with that state, and we will ask for the largest set of mutually commuting states, since we can then identify this with the {\em Cartan subalgebra} of the group involved.

The notion of a Cartan subalgebra (CSA) may not be familiar to all readers, but along with several other aspects of Lie group theory, will be explained in terms of a geometric algebra approach in the Lasenby (2022) paper in preparation. Suffice to say here, that the {\em roots} of the Lie algebra are composed of eigenvalues, one for each element of the CSA, and that the whole space can be viewed as a combination of the roots and the CSA elements. The number of the latter is known as the {\em rank} of the group, and the sum of the number of roots and the rank is the dimension of the overall space.

Now the octonionic left space has the following set of 7 states arising from chains which appear to be the maximal commuting subset:
\be
\begin{gathered}
T_1=e_1 \star \left( e_2 \star \left( e_4 \star \phi\right) \right), \quad T_2=e_1 \star \left( e_5 \star \left( e_6 \star \phi\right) \right), \quad T_3=e_2 \star \left( e_3 \star \left( e_5 \star \phi\right) \right), \quad T_4=e_3 \star \left( e_4 \star \left( e_6 \star \phi\right) \right),\\
 T_5=e_1\star\left(e_2 \star \left( e_3 \star \left( e_6 \star \phi\right) \right)\right), \quad T_6=e_1\star\left(e_3 \star \left( e_4 \star \left( e_5 \star \phi\right) \right)\right), \quad T_7=e_2\star\left(e_4 \star \left( e_5 \star \left( e_6 \star \phi\right) \right)\right) 
\end{gathered}
\label{eqn:Ti-defs}
\ee
Remarkably, and definitely exciting as regards the STA approach, when expressed in terms of STA operations, these 7 states are very simple:
\be
\begin{gathered}
T_1=-\gam_0\phi\gam_0, \quad T_2=\gam_1\phi\gam_1, \quad T_3=\gam_2\phi\gam_2, \quad T_4=\gam_3\phi\gam_3,\\
 T_5=\sig_3\phi\sig_3, \quad T_6=-\sig_2\phi\sig_2, \quad T_7=-\sig_1\phi\sig_1
\end{gathered}
\ee
We see that these are the 7 distinct reflections which can be carried out using unit STA elements. Note that we can use odd elements as well since reflection in either even or odd elements preserves evenness. We do not get extra states from using any versions including $I$, however, such as trying to reflect in $I\sig_i$ or $I\gam_i$ ($i=1,2,3$), since $I$ commutes with all even elements, and we will just get a change in sign, not a new state.

We can see immediately why all these states intercommute, since although we may get sign changes when two reflecting elements are interchanged, such as interchanging $\gam_0$ and $\sig_3$, which anticommute, we will get {\em two} such sign changes due to it happening each side of $\phi$, and overall the operations will commute.

It is very nice to be able to understand how these states, $T_j$, $j=1,\ldots,7$ are singled out, in number and type, by asking that they be due to reflections in unit STA elements, which brings a geometrical flavour to what is going on, which is certainly not captured in the assignments (\ref{eqn:Ti-defs}). It is also easy to verify that these operations leave the time component of the Dirac current invariant, as required for them to be equivalent to octonionic chains.

Carrying on with understanding the states in terms of a Lie group, we will assume that the $T_j$ form the Cartan subalgebra for the group of interest, and then use them, acting via a commutator product, to act on combinations of other states in the algebra in such a way as to give eigenvalues and eigenvectors. Specifically, we are looking for states which are simultaneously eigenvectors of all the $T_j$, $j=1,\ldots,7$, and the seven eigenvalues so found will give us the root vectors in an abstract 7d space corresponding to the seven $T_j$.

We will not give the details here, but we find that the 56 states left from the original 64 when we remove the identity and the seven $T_j$, split up into 7 groups of 8 states, each of which (i.e.\ the 8 elements of a given grouping) has non-zero eigenvalues with 4 of the $T_j$. The number of ways of choosing 4 things from 7 is of course 35, but one finds that each of the $T_j$ appears a total of 4 times, in the sense of having non-zero eigenvalues, and this limits the overall number of groupings to 7, as stated.

The net effect is to give us 56 root vectors in the abstract space of the $T_j$. Since the rank is 7 and the number of root vectors is 56, we initially believed that what we have got here is a particular version of the group $E_7$, plus one additional state, which is just the identity. The group $E_7$ is normally considered to have 133 dimensions and therefore 126 roots. However, it does also have a 56 dimensional representation (e.g.\ \cite{Brown+1969+79+102}), which is used for example in string theory\cite{hillmann2009generalized}. 
Having constructed the {\em Cartan matrix} for the roots we have found, it now appears that instead we are dealing with the group $SU(8)$, which indeed has 56 root vectors and 7 CSA vectors in the adjoint representation. This is still strongly linked to $E_7$, however, since it is a subgroup of $E_7$, and the two occur together jointly as part of the foundation of {\em supergravity theories} (see e.g.\ \cite{CREMMER1979141}).

More generally, it will be very interesting to see if the larger space of `STA states' referred to above allows a representation of the full $E_8$, and how this ties in with existing Geometric Algebra techniques for analysis of $E_8$ being used by Pierre Dechant\cite{dechant20178}, which have already shone an interesting and fresh light on the root structure of $E_8$. We should also mention the work of integrating the group $E_6$ with octonions, carried out over some years by Tevian Dray and Corrine Manogue (e.g.\ \cite{Manogue_2010}), and which is another example of explicitly seeking to integrate exceptional groups, particle physics and octonions, and for which we hope there will be interesting analogues within the STA.

\subsection{The group $G_2$ and the representation of quarks and gluons via the STA}

\label{sect:G2-and-quarks}

Turning to the other example we want to consider, the most direct way to approach this that we have found, starts from the 7 groups of 8 states each just discussed. Each group of 8 itself breaks into two groups of 4, distinguished by the mutual commutation of each of the 4 elements in a given split. One of these two subgroups (we are using `group' and `subgroup' in a non-technical way at this point, just to indicate a collection of objects) contains $e_i\star\phi$ and this happens for each of the 7 sets of 8, $i=1,\ldots,7$. We label the three remaining states in the $i$'th subgroup
\be
J_{1i}, \quad J_{2i}, \quad J_{3i}
\ee
and from what has already been said, all three mutually commute for a given $i$. Note calling these objects $J$'s is motivated by the fact that for $i=7$, then the differences between the three $J$'s match the differences between the $J_1$, $J_2$ and $J_3$ defined above in the context of the STA approach to $SU(3)$ in equation (\ref{eqn:hats-def}), i.e.\ we have
\be
J_{17}-J_{27}=\hat{J}_1-\hat{J}_2, \quad J_{27}-J_{37}=\hat{J}_2-\hat{J}_3, \quad J_{37}-J_{17}=\hat{J}_3-\hat{J}_1
\label{eqn:J-diffs}
\ee
This is important, since in the next step we indeed use differences between the $J$'s to define the quantities which are the generators of what we interpret as $G_2$. Specifically, we define 14 quantities $F_i$ and $M_i$ via
\be
F_i = \half\left(J_{1i}-J_{2i}\right), \quad M_i = {\textstyle \frac{\sqrt{3}}{6}}\left(\left(J_{1i}-J_{3i}\right)+\left(J_{2i}-J_{3i}\right)\right) = {\textstyle \frac{\sqrt{3}}{6}}\left(J_{1i}+J_{2i}-2J_{3i}\right)
\ee
Using equation (\ref{eqn:J-diffs}) in conjunction with the relationships to the Gell-Mann matrices given in equation (\ref{eqn:lambda-defs}), we can see immediately that we can identify $F_7$ with the generator $\lambda_3$ and $M_7$ with $\lambda_8$.

We will not go through the details here, but more generally, by comparing their STA effects, we find the identifications
\be
F_1 \leftrightarrow \lambda_7, \quad F_2 \leftrightarrow \lambda_5, \quad F_3 \leftrightarrow \lambda_6, \quad F_4 \leftrightarrow \lambda_2, \quad F_5 \leftrightarrow \lambda_1, \quad F_6 \leftrightarrow \lambda_4, \quad F_7 \leftrightarrow \lambda_3
\ee
and we have already said $M_7\leftrightarrow\lambda_8$. Thus we have all 8 $SU(3)$ generators present, along with another 6 new generators, the $M_i$, $i=1,\ldots,6$. The total 14 is the set of generators of the exceptional group $G_2$.

Now remarkably, although there is a very different starting point in terms of our 7 groups of 8 elements, the position we have just reached with $G_2$ is the same picture as already found by G\"unaydin and G\"ursey in 1973, when considering an octonionic model for quarks\cite{gunaydin1973quark}. Indeed, we have adopted their notation of $F$ and $M$ for the $SU(3)$ and (additional) $G_2$ generators, and they also use $J$ for the elements from which everything is assembled, though this choice was made independently in the current work.

However, as well as the starting point in terms of the organisation of states being different, we have the advantage here of being able to represent everything concretely in terms of the STA. We know this already for the $F_i$, since these are just the $\lambda$'s, and it is of interest to find expressions for the $J_{ki}$ themselves, since we can then form both the $F$'s and $M$'s from them via differences. To give an example, the $J_{k2}$, $k=1,2,3$ differences can be formed as follows:
\be
J_{12}-J_{22}=\phi I\sig_2 - I\sig_2 \phi, \quad J_{22}-J_{32}=I\sig_2 \left(\phi-\gam_0\phi\gam_0\right), \quad J_{32}-J_{12}=\phi I\sig_2 - I\sig_2 \gam_0\phi\gam_0
\ee
These are very simple, and are revealing in terms of how they involve $-I\sig_2$, which is the second octonion unit, and show how this quantity enters the second grouping of 8 states.

More generally, we can think of the $J_{ki}$ as corresponding to `rotations' amongst the unit octonions which will preserve the $i$'th one, whilst inducing rotations in pairs in the rest. It is in this sense $G_2$ arises as the {\em automorphism group} of the octonions. The restriction to {\em differences} of the $J$'s in doing this cuts out the 7 possible additional generators
\be
N_i = J_{1i}+J_{2i}+J_{3i}, \quad i=1,\ldots,7
\ee
in an exactly analogous way as $U(3)$ was restricted to $SU(3)$ by the condition on the $\alpha$'s given in equation (\ref{eqn:alpha-restriction}) above. As discussed in \cite{gunaydin1973quark}, the combination of the 14 $G_2$ generators and the additional $N_i$ generators just listed, yields 21 generators in total, which we can identify as the special orthogonal group of rotations in 7d space, $SO(7)$, which is known as a subgroup of $E_7$, tying in with our interpretation of what the full space of the left chains is.

To validate the identification of the group $G_2$, we need to show that as well as having 14 generators, it has rank two, and with a root diagram matching that expected. This brings us into contact with the approach used by Garrett Lisi\cite{Lisi:2007gv}.
Lisi does not consider octonions, and instead the important constructs are the Lie algebra elements treated in an abstract way, with their actions on one another being carried out via Lie brackets, and with the quantum numbers of the associated particles being `weights' within the root system of various exceptional Lie groups. We will only give brief details here, and will limit ourselves to just one example, but this will indicate how we can give a concrete version of this process, within the octonionic chain setup, and therefore also within the STA.

The example we use is the starting point of Lisi's study, which is about using $SU(3)$ to describe the gluons of the colour force, which is of course standard, but then as discussed in Section~2.1 of \cite{Lisi:2007gv}, using the 6 extra dimensions which $G_2$ introduces relative to the $SU(3)$ algebra which is embedded within it, to describe the quarks themselves. In particle physics notation, these are in a $\boldmath{3} + \bar{\boldmath{3}}$ `triplet' plus `anti-triplet' state.

Since everything (in this approach) is expressed in terms of the root vectors, we need to specify the Cartan subalgebra we are going to use. For this we need to have available all the commutation relations between our 14 generators (the $F_i$ and $M_i$ for $i=1,\ldots,7$), and we will list these in the more detailed publication to follow. However, this shows that the maximal size of a commuting subset is 2, which correctly corresponds to the rank of $G_2$. Here we will identify $F_7$ and $M_7$ as the CSA, since these work for both $SU(3)$ and the full $G_2$. So we need to find the simultaneous eigenvectors, using sums of the remaining 12 generators, and their asssociated eigenvalues (the Lie group `weights') under commutation with $F_7$ and $M_7$.

At this point we find a difference with what we are apparently able to do in term of forming combinations of states, and what Lisi, Furey and others are able to do, which relates to the question of whether we need to admit a commuting unit scalar imaginary into our system.

So far we have used only STA elements in everything we have done, apart from the first discussion of $SU(3)$ in Section~\ref{sect:first-su3}. The STA does not include a scalar imaginary $i$, but we have nevertheless been able to reach a long way into study of the octonionic chains, and also the groups $E_7$ and $G_2$, and this is perhaps remarkable in itself.

Other authors, e.g. Furey\cite{Furey:2015tqa,Furey:2018yyy}, bring in $i$ at the very beginning of their work, for example in the definitions of the $SU(3)$ elements, and Lisi brings in $i$ to his work at this point in his Section~2.1 in forming `complex' combinations of $SU(3)$ elements, of the form (in Gell-Mann matrix terms) $\lambda_2\pm i \lambda_1$, $\lambda_5\pm i \lambda_4$ and $\lambda_7\pm i \lambda_6$. The point about these combinations (6 of them), is that they are able to be simultaneous eigenvectors of the matrix equivalents of our two CSA elements, namely of $\lambda_3$ and $\lambda_8$. Without forming complex combinations, and treating each generator individually, one finds that one in any case needs imaginary eigenvalues, so again an $i$ has to be apparently introduced.

We can avoid this to some extent in our approach, by working with {\em ordered pairs} of states, which we can conveniently represent via a row vector, so for example, the 6 states just described would be
\be
\begin{pmatrix} F_4, & \pm F_5 \end{pmatrix}, \quad \begin{pmatrix} F_2, & \pm F_6 \end{pmatrix}, \quad \begin{pmatrix} F_1, & \pm F_3 \end{pmatrix}
\label{eqn:six-gluon-states}
\ee

Taking the simplest example, let us consider the action of the CSA elements $F_7$ and $M_7$ on the first of these pairs of states. Writing e.g.\
\be
F_7 \begin{pmatrix} F_4, & F_5 \end{pmatrix} \quad \text{to mean}  \quad \half\begin{pmatrix}  [F_7,F_4], & [F_7,F_5] \end{pmatrix}
\ee
where $[\ ,\ ]$ is the commutator, then we find
\be
F_7 \begin{pmatrix} F_4, & F_5 \end{pmatrix} = \begin{pmatrix} F_4, & F_5 \end{pmatrix}\begin{pmatrix} 0 & -1 \\ 1 & 0 \end{pmatrix} , \quad F_7 \begin{pmatrix} F_4, & -F_5 \end{pmatrix} =- \begin{pmatrix} F_4, & -F_5 \end{pmatrix}\begin{pmatrix} 0 & -1\\  1 & 0 \end{pmatrix} , \quad \text{and} \quad M_7 \begin{pmatrix} F_4, & \pm F_5 \end{pmatrix} =0
\label{eqn:eigenvec-example}
\ee
This translates through, in Garrett Lisi's terms, to the pair $\begin{pmatrix} F_4, & \pm F_5 \end{pmatrix}$ being simultaneous eigenvectors of $\lambda_3$ and $\lambda_8$ (our $F_7$ and $M_7$ respectively), with weights in the 2d root space of $(1,0)$ and $(-1,0)$. To state this further in a particle context, we note that Lisi would identify $\begin{pmatrix} F_4, & F_5 \end{pmatrix}$ as an anti-red green gluon and $\begin{pmatrix} F_4, & -F_5 \end{pmatrix}$ as a red anti-green gluon. We can then continue with the other two pairs of states in (\ref{eqn:six-gluon-states}) and similarly, via equations like (\ref{eqn:eigenvec-example}), find their root vector weights in the 2d CSA of $F_7$ and $M_7$, and these match exactly with those found for the various gluon states in Lisi.

While implemented in a novel fashion, what we are looking at here is at an abstract level just the usual aspects of $SU(3)$ in relation to the colour force. What Lisi does next is to suggest identifying the remaining 6 generators of $G_2$ with the quark states themselves. He does not give explicit details of this identification, but just works out the weights of the corresponding root vectors in the CSA space.

In our terms, we can give an explicit version of this as follows. We define three quark states, $q_1$, $q_2$, $q_3$, along with three anti-quark states, $\bar{q}_1$, $\bar{q}_2$, $\bar{q}_3$ via the ordered pairs
\be
q_1 = \begin{pmatrix} M_3, & M_1 \end{pmatrix}, \quad \bar{q}_1 = \begin{pmatrix} M_3, & -M_1 \end{pmatrix}, \quad q_2 = \begin{pmatrix} M_2, & -M_6 \end{pmatrix}, \quad \bar{q}_2 = \begin{pmatrix} M_2, & M_6 \end{pmatrix}, \quad q_3 = \begin{pmatrix} M_5, & -M_4 \end{pmatrix}, \quad \bar{q}_3 = \begin{pmatrix} M_5, & M_4 \end{pmatrix}
\label{eqn:quark-states}
\ee

The interactions of the gluons with the quarks are then represented by the actions of the $F_1$, \ldots, $F_7$ and $M_7$ $SU(3)$ generators with these quark states, carried out as usual by commutators. The particular combinations in (\ref{eqn:quark-states}) are each eigenvectors of what we can call the `hypercharge' operator, which in the current context turns out to be the operator
\be
Q=\frac{1}{2}\left(\frac{1}{\sqrt{3}}M_7 + F_7\right)
\ee

Using the multiplication properties of the $F$'s and $M$'s, we find the corresponding eigenvalues are (respectively) $+2/3$ and $-2/3$ for $q_1$ and $\bar{q}_1$; $-1/3$ and $+1/3$ for $q_2$ and $\bar{q}_2$ and $1/3$ and $-1/3$ for for $q_3$ and $\bar{q}_3$, which fits in well with their identification as quark states. Note looking at the eigenvalues w.r.t.\ $F_7$ and $M_7$ individually, we find that these line up precisely to give the remaining 6 roots in the root diagram of $G_2$, as stated by Lisi, confirming that in the STA version we are indeed working with $G_2$.

Additionally we can consider the action of the remaining $F$'s on these 6 $M$ states. We find that the combinations given in (\ref{eqn:six-gluon-states}), identified as gluon states, move the quarks around within the three $q_i$ and within the three $\bar{q_i}$ states, though do not mix them. This again corresponds well to what we would expect for a `triplet' and `anti-triplet' set of quark states being acted upon, so overall we have a concrete representation of the gluon and quark states, and their interactions, all expressible within the STA. However, as we have seen, this is modulo questions about whether we in fact need a commutative scalar imaginary $i$ to make this work.

Addressing this, it might be possible to argue that equation (\ref{eqn:eigenvec-example}) is just a way of organising our knowledge about the algebraic relations satisfied by the $F_i$ and $M_i$, which have all been worked out (ultimately in the STA), without any importation of a scalar unit imaginary $i$. This is a possible way of thinking, but on the other hand, one way of introducing complex numbers is precisely via introducing ordered pairs of real numbers, along with a way of flipping between the two states of a pair, as in the matrix $\begin{pmatrix} 0 & -1 \\ 1 & 0 \end{pmatrix}$. Thus it could be argued that we do indeed need to introduce what is effectively a `tensor product' of the octonionic chain states with the complex numbers $\mathbb{C}$, precisely as hypothesised at the beginning of e.g.\ the work by Cohl Furey. We will return to this question in the more detailed work to follow, and propose a slightly different resolution of the question there. In the meantime, we reiterate that it is only for the `packaging' of the octonionic STA states into pairs that we currently need constructions like (\ref{eqn:eigenvec-example}), and this has happened right at the end of our developments, rather than being an intrinsic part of the beginning.

\section{Future prospects}

So far, in particle terms, we have only considered a limited segment of the Standard Model, and there is much still to do in terms of having a complete version expressible in the STA. An interesting feature concerning this is that recently, following Furey's work,  Gording and Schmidt-May\cite{gording2020unified} report that they have fitted the entire SM, with all three generations, into the 64-dimensional octonionic left multiplication space. This is carried out, however, using a complex $8\times 8$ matrix approach, rather than using either Clifford algebras or octonions, and so is not immediately interpretable in the current context. Nevertheless it does bode well for the basic enterprise here.

A further interesting avenue of exploration is the significance of the 7 sets of three J's in Section~\ref{sect:G2-and-quarks}. Since these mutually commute within a set, then the $F$, $M$ and $N$ generators formed out of them can all be easily exponentiated to yield finite states within $SU(3)$, $G_2$ and $SO(7)$. In the context of the $Cl(0,6)$ space equivalent to the octonionic left multiplcation space, these $J$'s are not all bivectors, but it will be interesting to relate these to the decomposition of the bivector generators of $SU(3)$ into commuting blades in the Roelfs and de Keninck approach\cite{roelfs2021geometric,roelfs2021graded}. They will possibly also provide a route to the same type of decomposition in the larger groups $G_2$ and $SO(7)$.

In relation to Lisi's hope of using $E_8$ to provide a unified model for all the symmetries of both particle physics and gravitation, we have only looked at one corner of this. Although that worked well (see Section~\ref{sect:G2-and-quarks}), the basic enterprise of including gravity in the same type of Lie group connection as that for the strong and electroweak forces looks problematic in our approach, since Gauge Theory Gravity\cite{1998RSPTA.356..487L} requires an $h$-function to express `position gauge' invariance, whereas this type of object is not present in the Lisi formulation. This remains to be understood further, but in the meantime it will certainly be of interest to look at the other non-gravitational aspects of Lisi's $E_8$ proposals.

There is a good deal more to do on the issue of the $\psi(x) \mapsto \psi(B,x)$ construction, in which the Dirac wavefunction becomes a linear function of a bivector. The transition to octonionic transformations of a full Dirac spinor $\phi$, rather than just its bivector part $B$, means that we will now need to consider Dirac wavefunctions (of the type that respond to real, external, spacetime Lorentz transformations), as being linear functions of a full `internal' spinor, not just an internal bivector $B$, i.e.\ we will need a construction $\psi(x) \mapsto \psi(\phi,x)$ instead of just $\psi(x) \mapsto \psi(B,x)$. This is in some sense a fairly small step, since dependence on just two new components is introduced, but it looks even more unfamiliar, and the consequences need to be worked out.

Finally, we have a definite issue left over, with the role of a commutative scalar imaginary $i$, and whether this is necessary or not. In the work of Dixon (e.g.\ \cite{dixon2004division}), a tensor product of the octonionic states is taken not just with the complex numbers, but with the quaternions as well, leading to what is called the RCHO algebra. Whether either of these steps (introduction of $\mathbb{C}$ or $\mathbb{H}$) is going to be necessary in our approach is something still to be determined.

Overall, we are still left with a question of which are the important entities as far as Nature is concerned: are they the STA, the Octonions, the larger Clifford spaces to which the octonionic chains are related, or the abstract Lie groups, or perhaps some combination of them?  In any case, with regard to the Octonions, we hope the present work aids people in the GA community becoming more familiar with them, and in a context where they can easily carry out the computations, and preserve a sense of an underlying geometrical picture.

\bibliography{sig_supplementary_references}

\end{document}